# Single shot, double differential spectral measurements of inverse Compton scattering in linear and nonlinear regimes

Y. Sakai[1], I. Gadjev[1], P. Hoang[1], N. Majernik[1], A. Nause[1], A. Fukusawa[1], O. Williams[1], M. Fedurin[2], B. Malone[2], C. Swinson[2], K. Kusche[2], M. Polyanski[2], M. Babzien[2], M. Montemagno[2], Z. Zhong[2], P. Siddons[2], I. Pogorelsky[2], V. Yakimenko[2,5], T. Kumita[3], Y. Kamiya[4], J. B. Rosenzweig[1]

[1]UCLA Department of Physics and Astronomy, 405 Hilgard Ave., Los Angeles, CA 90095
[2]Brookhaven National Laboratory, Upton, NY 11973
[3]Tokyo Metropolitan University, Tokyo 192-0397, Japan
[4]ICEPP, Tokyo University, Tokyo 113-0033, Japan
[5]SLAC National Accelerator Laboratory, Menlo Park, CA 94025,

Inverse Compton scattering (ICS) is a unique mechanism for producing fast pulses — picosecond and below — of bright X- to $\gamma$-rays. These nominally narrow spectral bandwidth electromagnetic radiation pulses are efficiently produced in the interaction between intense, well-focused electron and laser beams. The spectral characteristics of such sources are affected by many experimental parameters, such as the bandwidth of the laser, and the angles of both the electrons and laser photons at collision. The laser field amplitude induces harmonic generation and importantly, for the present work, nonlinear red shifting, both of which dilute the spectral brightness of the radiation. As the applications enabled by this source often depend sensitively on its spectra, it is critical to resolve the details of the wavelength and angular distribution obtained from ICS collisions. With this motivation, we present here an experimental study that greatly improves on previous spectral measurement methods based on X-ray $K$-edge filters, by implementing a multi-layer bent-crystal X-ray spectrometer. In tandem with a collimating slit, this method reveals a projection of the double-differential angular-wavelength spectrum of the ICS radiation in a single shot. The measurements enabled by this diagnostic illustrate the combined off-axis and nonlinear-field-induced red shifting in the ICS emission process. They reveal in detail the strength of the normalized laser vector potential, and provide a non-destructive measure of the temporal and spatial electron-laser beam overlap.

**INTRODUCTION**

Inverse Compton scattering (ICS) [1,2,3] is an emerging technique for obtaining narrow bandwidth, highly directional X-rays from the collision of intense relativistic electron beams and lasers. The radiation characteristics of such ICS sources are similar to those obtained from undulators in storage ring-based light sources and the spontaneous radiation from free-electron lasers (FELs), with a qualitative difference. This is found in the scale of the periodic fields that provoke the radiation; in contrast to the centimeter-period $\lambda_u$ associated with a magnetostatic undulator array, in ICS one employs a counter-propagating electromagnetic wave that gives an equivalent period of $\sim \lambda_L / 2$, where $\lambda_L$ is the (optical-to-infrared) laser wavelength. As in both the ICS and magnetic undulator cases the back-scattered radiation is Doppler-shifted upwards in frequency by a factor $2\gamma^2$, where $\gamma$ is the electron Lorentz factor, the use of a "laser undulator" provides a path to obtaining very energetic photons, ranging from the keV to MeV using only modest energy, 20 MeV to few-100 MeV electron linear accelerators (linacs). With smaller energies demanded, an ICS instrument is potentially very compact in size, permitting X-ray light source research facilities in small university-scale laboratories [4]. Emerging applications in medicine [5,6] that can benefit from nearly monochromatic hard (~10-100 keV) X-rays are found in both diagnosis — *e.g.* in phase contrast or dual-energy digital subtraction imaging — and therapy, where *K*-edge absorption may be used to greatly enhance local X-ray dose (*e.g.* photon activation therapy [7]). The potential to reach sub-picosecond X-ray pulses further opens up new possibilities in ultra-fast experiments, particularly in single-shot, pump-probe investigations [8,9]. At higher electron beam energies, in the 100's of MeV range [10], one may produce MeV photons for fundamental nuclear photonics experiments [11], or for detection of special nuclear materials [12].

The robustness of all the applications listed above depends, to varying degrees, on the details of the ICS spectral distribution in wavelength and angle. For a single electron radiating under the influence of a counter-propagating plane wave, this spectral dependence of scattered radiation wavelengths $\lambda_r$ on electron energy and emission angle can be approximately written as

$$\lambda_r = \frac{\lambda_L}{4\gamma^2}\left[1+(\gamma\theta)^2 + \tfrac{1}{2}a_L^2\right]. \tag{1}$$

Here $\theta$ is the emission angle measured from the electron propagation direction, and the normalized vector potential, $a_L = eE_L\lambda_L/2\pi m_e c^2$, (with $E_L$ the laser's electric field amplitude) measures the degree to which the electromagnetic wave induces relativistic transverse motion in the oncoming electron. This transverse motion has the effect of lowering the effective $\gamma$ of the directed longitudinal motion, and is thus sometimes referred to as the *mass shift* effect [13]; it is also commonly known as the nonlinear red shift. In the present work, we indicate the maximum value of $a_L$ that is associated with the peak value of $E_L$ in the laser beam as $a_0$.

There are finite collision-angle effects due to deviations of both the electron and laser photon trajectories [14] from the axial direction that increase the scattered bandwidth, and reduce the associated brightness. Thus highly focused systems, while producing more photons in the collision, yield a larger spread in photon energy at a given angle. A further, fundamental constraint on source brightness arises when one focuses the laser to higher intensity, again producing more scattered photons, but concomitantly increasing the value of $a_L$. A value of $a_L$ not small compared to unity causes strong spreading, proportional to $a_L^2$, of the emitting ICS spectrum. An additional spectral spreading effect occurs when one approaches or exceeds $a_L \sim 1$, as harmonics of the radiation are produced [15]. This phenomenon again serves to lower the spectral brightness of the ICS radiation produced. As such, both nonlinear red shift and harmonic generation have attracted recent experimental attention.

## EXPERIMENTAL DESCRIPTION

Pursuant to the above described physics motivations, a recent experimental study has examined the relevant aspects of the nonlinear ICS interaction [16], with cases where $a_L \sim 0.6$ showing strong evidence of both red shifting, as well as harmonic scattering up to third order. These experiments [17,18] were performed employing metallic foils that exploit *K*-edge absorption to produce low-pass and band-pass filters. While these relatively crude filters adequately permit isolation of harmonics and sensitive determination of the threshold for nonlinear red shifting, they do not yield detailed information on the spectrum. Further, the correlations between angle and wavelength may only be approximately determined. As an experimental understanding of the details of the scattered spectrum is essential for applications of ICS, a more powerful diagnostic, simultaneously revealing both angular and energy spectral information is demanded. This is particularly urgent in the experimental scenario discussed below, where improvements in the laser system quality, in particular an increase of peak power from 0.4 TW to 0.8 TW, yield an unprecedented value of the normalized vector potential, with $a_L$ reaching near unity — or a factor of over 2.5 increase in nonlinear redshift. This context yields a rich array of observable interaction physics that can be quantified by more insightful diagnostic techniques.

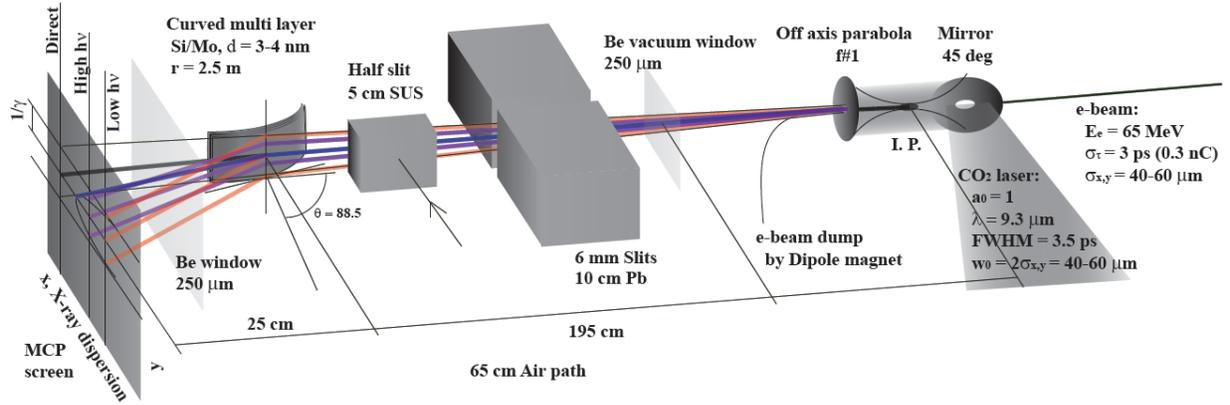

Figure 1. Schematic layout of multi-layer bent multi-layer spectrometer used to obtain double differential ICS spectrum in BNL ATF experiments

As such, we have developed and utilized in ICS experiments in the strongly nonlinear regime an X-ray spectrometer [19] capable of revealing the double differential radiation spectrum (DDS) obtained from the scattered photons. This X-ray spectrometer, which is shown schematically in Figure 1, consists of a collimating slit array, a bent molybdenum-silicon (Mo-Si) multilayer crystal to disperse the wavelength spectrum, and a micro-channel plate (PHOTONIS MCP 40/12/10/8 I 60:1 EDR KBR 6" FM P46), in combination with a phosphor screen and a 12-bit CCD camera (Basler, SCA1400-17). The MCP and its readout form a two-dimensional position sensitive detector optimized to produce X-ray images. The bent multilayer crystal itself is constructed with a periodic Bragg structure having 45 Mo-Si pairs, each layer of thickness ~20 Å. It reflects X-rays in the same manner as a natural crystal with the lattice spacing of ~40 Å, with the relation between the reflection angle and X-ray energy given by the Bragg's law, indicated here as:

$$\sin(\theta_B) = \frac{0.155 m}{E_{X-ray}(\text{keV})}, \qquad [2]$$

where $\theta_B$ is the Bragg angle and *m* is the reflection order.

The multilayer crystal is curved into a cylindrical section [19]. The curvature permits the incidence angle of the ICS X-rays impinging on the crystal to vary from 25 mrad to 10 mrad, corresponding to the X-ray energies ranging from (Eq. 1) 6 keV to 15 keV. This energy range overlaps optimally with the expected fundamental component of the X-ray spectrum in the BNL ATF ICS source. These X-ray energies are dispersed along the (horizontal) *x*-direction in Figure 1, while in the orthogonal direction the mapping of reflection angle along the *y*-axis directly represents the vertical emission angle of X-rays from the collision point. In order to separate the angular emission from the wavelength dispersion information, previous to encountering the bent crystal the X-rays pass through a narrow (in the *x*-direction) slit. Thus the X-ray distribution detected and recorded at the MCP and associated apparatus closely approximates a section of double differential spectrum, $U(\lambda, \theta_y)$. An example of such a DDS section is shown below in Figure 3. In the current case, where we examine the fundamental ICS harmonic generated with a linearly polarized laser, the angular spectrum is not exactly axially symmetric for large $a_0$, being elongated along the direction of the polarization. Thus the section of the DDS $U(\lambda, \theta_y)|_{\theta_x=0}$ is marginally dependent on the relative direction of the laser polarization and the slit direction; in the experiments reported here the slit and polarization directions are at 30 degrees with respect to each other. For the following discussion, which concentrates mainly on the near-axis component to analyze nonlinear redshift effects, this polarization direction does not affect the spectrum.

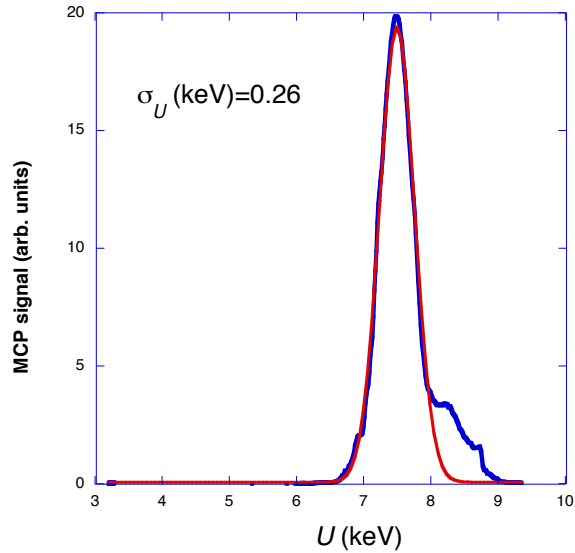

Figure 2. Results of calibration tests of spectrometer at BNL NSLS-I X15A (in blue), measured dispersion (photon energy vs. radiation angle *x*), displaying a resolution at 7.5 keV central energy of 0.26 keV (rms), as illustrated by Gaussian fit to peak (in red).

The calibration of the mapping of dispersive angle to X-ray energy in this instrument was accomplished using known-energy distributions of photons at a number of different energies obtained from the NSLS X-ray light source beamline X15A, using a Si 111 and 333 monochromator having resolution $\Delta v/v = 2E-4$, and a hard X-ray CCD BAS2500 image plate reader. The rms resolution of the bent crystal spectrometer deduced from this calibration exercise is 0.26 keV at a central energy of 7.5 keV, in the middle of the spectral band of interest. We note that the fractional bandwidth is approximately the inverse of the number of Bragg layers, as expected.

To create the scattered photons, a similar set of experimental conditions was used at the BNL ATF ICS source as were employed in the experiments of Ref. 15, with the notable exception of an increase in laser peak power. In the experiments reported here the electron beam parameters are: ~3 ps rms pulse length, ~1.3 mm-mrad normalized emittance, 0.3 nC total charge, and $E_0$=65 MeV. This beam is generated by a RF photoinjector and subsequent linac, and subsequently transported and focused to ~$\sigma_e$=40 μm rms spot where it is collided with a ~3.5 ps FWHM $CO_2$ laser [20] ($\lambda_L = 9.3$ μm). We note that the electron beam focus is sufficiently relaxed that, given the emittance, the induced bandwidth from electron angles during collision is small, as discussed further below. This laser pulse has between 1.5 and 3.0 J of total energy, corresponding to ~0.4-0.8 TW peak power, and is $w_0$=2$\sigma_L \approx 60$ μm, normalized vector potentials reaching up to $a_0$=1 are obtained. The electrons are brought into collision with the laser by use of strong quadrupole focusing magnet and an alignment pinhole. The electron beam rms size at focus is significant in determining the spectral shape of the ICS radiation, as the condition $\sigma_e \approx \sigma_L$ produces a notable experimental spectral signature, discussed below. Temporal synchronization of the electrons and laser is measured and optimized via use of a Ge semiconductor switch [21] located at the laser- and electron-beam interaction point. The ICS X-rays are extracted through a 250 μm-thick Be window and 0.5 m air path before passing through the bent multi-layer spectrometer.

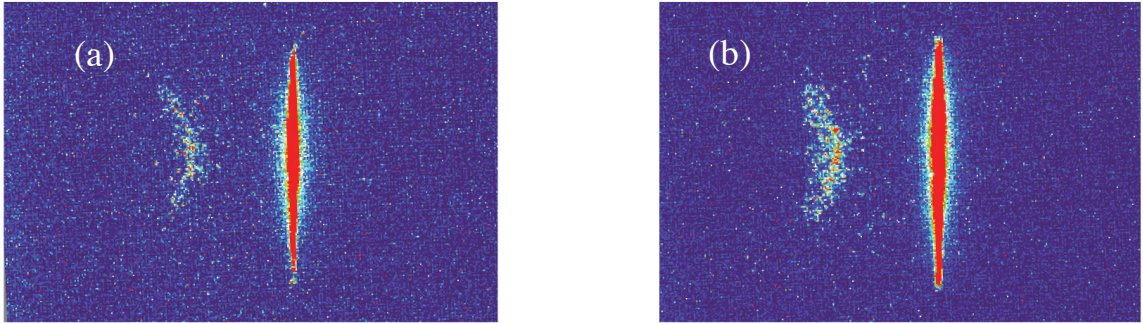

Figure 3. (a) ICS photon distribution for 1.5 J laser interaction ($a_L$=0.7), after passage through bent crystal, showing both undiffracted (vertically oriented bright red spot) and diffracted (dispersed) component; (b) higher laser energy case, with 3.0 J and $a_0$=1. Note the curved shape of the diffracted double-differential spectrum in both images, as well as the enhanced width due to nonlinear red-shifting in the higher laser energy case.

## RESULTS AND DISCUSSION

Two images of the ICS X-rays obtained after the spectrometer and slit apparatus are shown in Figure 3, corresponding to the partially and strongly nonlinear cases with 1.5 J ($a_0$=0.7) and 3.0 J ($a_0$=1) laser, respectively. In these images the main features of the projected ICS DDS $U(\lambda,\theta_y)$ are displayed. First, there is a curved intensity distribution corresponding to the parabolic off-axis red shifting indicated by the $\theta$-dependence shown in Equation 1. To illustrate this dependence, we show in Figure 4 the spectral edges (15% integrated intensity from the minimum and maximum energy limits of the observed X-ray distributions, respectively) as a function of emission angle $\theta_y$, corresponding to the $a_0=1$ laser case in Figure 3. As can be seen from the data and the predictions of the particle-tracking-based, Lenard-Wiechert (L-W) simulations [16] of the experimental scenario, utilizing Gaussian laser intensity and electron beam density profiles, whose physical significance is examined further below, one observes a very large bandwidth spectrum at a given $\theta_y$, even for the less energetic, lower $a_0$ case. The details of the wavelength spectral shape at constant $\theta_y$ revealed in this DDS image are central to the current investigation, and are analyzed further in the discussion following.

In order to appreciate the physics processes involved in the creation of the wavelength spectral distribution shape due to the intense-laser induced nonlinear red (mass shift) shifting effect. some theoretical discussion is warranted. We begin by examining the broadening of the spectral distribution at a given $\theta_y$ due to the nonlinear red shifting process. We concentrate on the local changes in amplitude of the electric field in the laser and their effects on the nonlinear red shift, for the moment de-emphasizing the fine structure introduced in the spectral shape by interference effects [15,22,23] that arise from the time-dependence of the emission process [24]. Such effects are difficult to resolve in the experiments with the present spectrometer, a point we return to later.

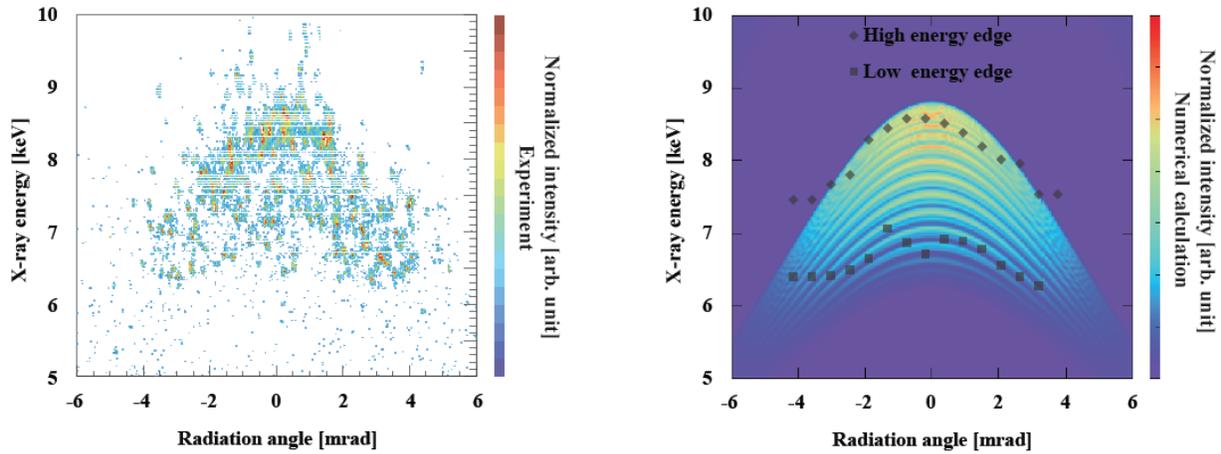

Figure 4. (left) Calibrated ICS DDS data through 2X 250 μm Be window and 65cm air path, 3 J laser energy case, using bent crystal system, from Fig. 3. (right) Analysis of angularly resolved energy distribution of ICS X-rays from measurement. For a given $\theta_y$, high energy and low energy edges (15% integrated intensity) are shown. Predictions of the Lenard-Wiechert calculation are shown in color map.

Theoretically, this model is equivalent to assuming that both the local Compton scattering probability, and nonlinear red shifting of the scattered photon produced with this probability, are proportional to the laser intensity, $a_L^2 \propto I$. This approximation ignores the wave-phase derived effects observed in the Lenard-Wiechert model and in more sophisticated theoretical treatments [24]. We assume for simplicity that the distribution of laser intensity is Gaussian both in the temporal and radial dimensions; likewise we also assume that the electron beam has a Gaussian transverse distribution of rms width $\sigma_e$. In this analysis, we also assume that the electron transverse position does not notably change during the laser-electron collision. This condition is well obeyed, as the minimum beam $\beta$-function, which measures the transverse spreading scale length of the electron beam, is ~4 cm; this is much longer than the interaction length. We also assume that we can ignore diffraction effects in the laser focus during collision, thus requiring (for Gaussian temporal pulse profiles) $Z_R > cT_L/4.7$. Here we have a laser Rayleigh range $Z_R$=0.5 mm, and spatial pulse length (FWHM) $cT_L$=0.22 mm, and the approximation holds well.

The structure of the ICS spectral distribution indeed arises in part from the time variation in the intensity, parameterized by $a_L^2(t)$, which as implied above is viewed as a property of the laser alone in the case of head on laser-electron beam collisions. The ICS spectral shape is also due to the transverse variation of the laser field. The degree to which this effect asserts itself is dependent on the parameter $\kappa \equiv \sigma_L / \sigma_e$, the ratio of the electron beam rms transverse spot to that of the laser beam. For $\kappa \gg 1$, the transverse variation in the field is not significant, but we shall see that for $\kappa \sim 1$ or below, transverse field variations are notable and the ICS spectral shape is strongly affected. This is indeed the case for the present experiments, and thus the measurement of the spectral shape at constant $\theta_y$ permits experimental probing of the overlap between the electron and laser beam distribution, *i.e.* an independent, nondestructive determination of $\kappa$.

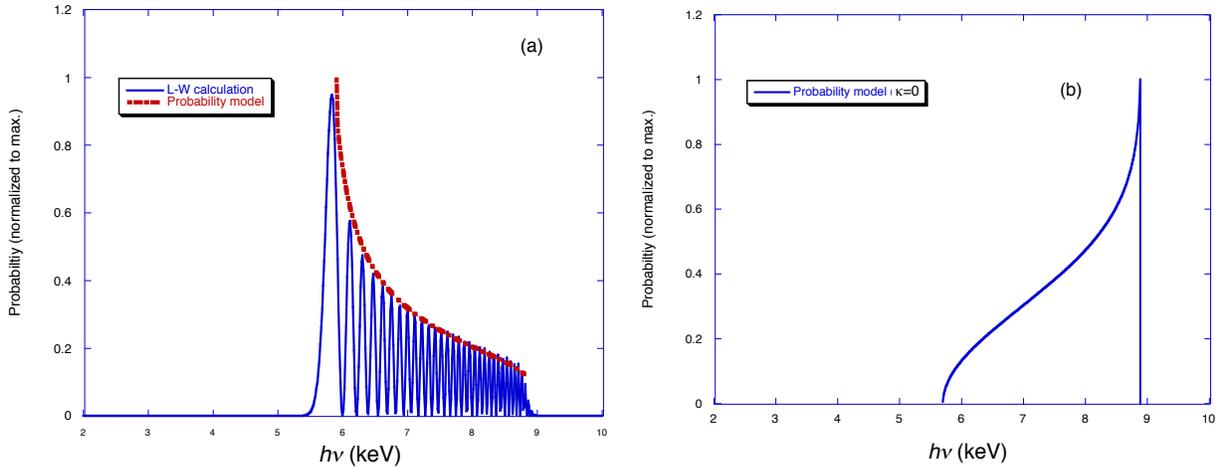

Figure 5. (a) ICS spectrum considering only longitudinal field variation (maximum photon energy $h\nu_{max}$=8.7 keV, $a_0 = 1$, maximum redshift 2.9 keV) from probability model (red dash). Also shown, the ICS spectrum as calculated by L-W model, which displays self-interference derived oscillations in this spectral intensity. (b) The prediction of the probability model when the transverse field variation is included ($\kappa=0$, with $h\nu_{max}$=8.7 keV, $a_0 = 1$), showing peak in spectrum.

In the case of a temporal and transverse spatial, axially symmetric Gaussian profile, with the assumed intensity distribution

$$I(r,t) \cong I_0 \exp\left[-\frac{r^2}{2\sigma_L^2}\right]\exp\left[-\frac{t^2}{2\sigma_t^2}\right], \qquad [3]$$

we can obtain analytical estimates for the spectral shape that yield information on the spatial-temporal characteristics of the interaction.

Concentrating first on the temporal dependence — assuming no transverse variation in the intensity, nor any other contributions to *homogeneous* spectral broadening (*i.e.*, energy spread, electron or laser photon angles) — the on-axis ($\theta_y = \theta_x = 0$) spectral shape is described, following Eq. 1, by the distribution

$$P_1(\Delta\lambda) = \frac{C_1}{\sqrt{\ln(\Delta\lambda_{max}/\Delta\lambda)}}, \qquad [4]$$

where $\Delta\lambda$ is the nonlinear red shifting, $C_1$ is a normalization constant, and the distribution function is weakly divergent near the maximum nonlinear redshift, $\Delta\lambda_{max} = \lambda_L a_0^2 / 8\gamma^2$. This divergence of course is eliminated in practice by homogeneous broadening effects. The predictions of this straightforward, heuristic scattering model are consistent with the results of a full Lenard-Wiechert (L-W) calculation [16], as displayed in Figure 5, which uses the maximum photon energy and redshift parameters encountered in the scenario of Figures 3(b) and 4. The spectral shape is seen to be reproduced well, missing only the oscillations due to the self-interference effects described in Refs. 15 and 22.

This strong peak near maximum nonlinear red shifting can be eliminated through the effects of the transverse variation of the laser intensity. To see this we first analyze the scattering probability associated with the change in laser intensity off-axis, but now ignoring for the moment the longitudinal (temporal) variation of the field. In this analysis, we take into account the effect of the weighting of the probability with radius *r* as well as the possibility of differing beam sizes, *i.e.* different values of $\kappa$, which serves as a parameter classifying the spectral shape. This analysis yields another surprisingly simple relation,

$$P_2(\Delta\lambda) = C_2 \left(\frac{\Delta\lambda}{\Delta\lambda_{max}}\right)^{\kappa-1}, \qquad [5]$$

with $C_2$ also indicating a normalization constant. Note that for $\kappa>1$, the probability distribution is again found to favor large nonlinear redshift $\Delta\lambda$, but for small laser beam sizes $\kappa<1$, smaller redshifts are more probable. For $\kappa=0$, the probability distribution is dominated by the effect of transverse variation in laser intensity; for $\kappa\gg1$, Eq. 5 approaches a $\delta$-function at $\Delta\lambda = \Delta\lambda_{max}$.

We wish to include both longitudinal and transverse effects to find a probability of observing a certain redshift $\Delta\lambda$. To do so we must convolve Eqs. 4 and 5, treating the total probability associated with Eq. 5 as yielding the maximum local redshift $\Delta\tilde{\lambda}$ given by all elements of the statistical ensemble indicated by differing *r* values. This convolution is written as

$$P_t(\Delta\lambda) = C_t \int_{\Delta\lambda}^{\Delta\lambda_{max}} \left(\frac{\Delta\tilde{\lambda}}{\Delta\lambda_{max}}\right)^{\kappa-1} \frac{1}{\sqrt{\ln[\Delta\tilde{\lambda}/\Delta\lambda]}} d(\Delta\tilde{\lambda})$$

$$= \frac{2C_t}{\Delta\lambda_{max}} \left(\frac{\Delta\lambda}{\Delta\lambda_{max}}\right)^{\kappa} \int_0^{\sqrt{\ln(\Delta\lambda_{max}/\Delta\lambda)}} \exp(\kappa u^2) du = \frac{2C_t}{\Delta\lambda_{max}} \left(\frac{\Delta\lambda}{\Delta\lambda_{max}}\right)^{\kappa} \sqrt{\frac{\pi}{4\kappa}} \mathrm{erfi}\left(\sqrt{\kappa \ln\left(\frac{\Delta\lambda_{max}}{\Delta\lambda}\right)}\right), \qquad [6]$$

with $C_t$ again indicating a normalization constant. To illustrate the limit of a very narrow laser beam, where the field variation in the transverse dimension is maximized, we examine the case $\kappa \rightarrow 0$, and obtain the asymptotic expression

$$P_t(\Delta\lambda)_{\kappa=0} = \frac{2C_t}{\Delta\lambda_{max}} \sqrt{\ln\left(\frac{\Delta\lambda_{max}}{\Delta\lambda}\right)}, \qquad [7]$$

which gives an opposite spectral dependence compared to that of Eq. 4, peaked at small $\Delta\lambda$. This behavior is shown in Figure 5(b). For values of $\kappa$ near unity, as is the case for the present experiments, a transition between the two limiting behaviors occurs. The spectral peak is broader for such cases, and found at an intermediate point between very low and maximum $\Delta\lambda$, as can be seen from exploration of Eq. 6. As can be ascertained from Figure 6, the convolution leaves the highest redshift probabilities unchanged as $\kappa$ is varied, while the probability of having a lower redshift goes down as the laser size increases relative to that of the electron beam, eventually reaching the result obtained in Eq. 4. The comparison between probability and L-W models has been checked, and the found to be in excellent agreement for values of $\kappa$ ranging from 0.1 to 40.

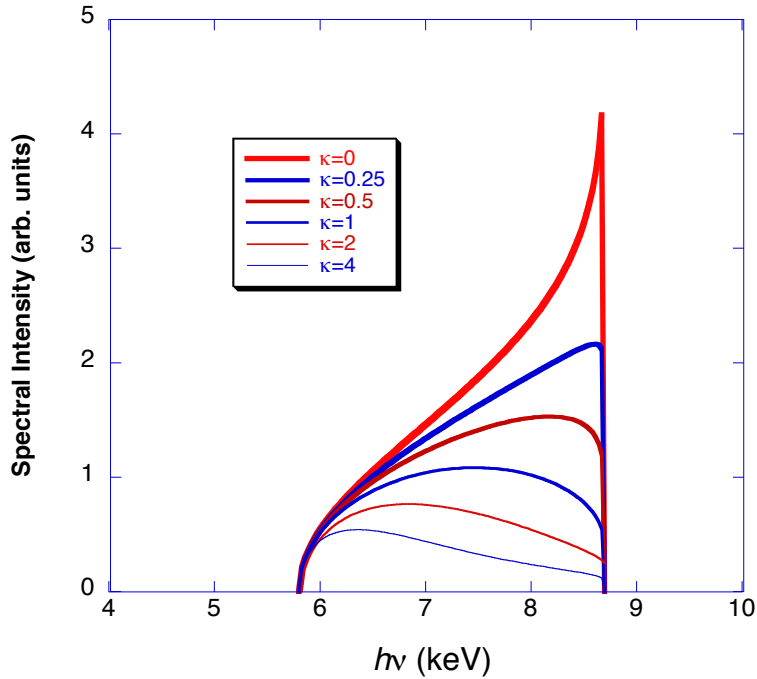

Figure 6. ICS spectrum with maximum photon energy $h\nu_{max}$=8.7 keV, $a_0 = 1$ (maximum redshift 2.9 keV) from probability model convolution of Eq. 6, for varying ratios of laser to electron beam size: $\kappa$=0, 0.25, 0.5, 1, 2, and 4.

We are now in a position to examine the near-axis spectrum obtained from the maximum-intensity laser case obtained from the data given in Figure 3(b). These data are obtained from the images in Figure 3(b) using the established calibration between angle and photon energy $h\nu$. A nearly on-axis section of the DDS having a width in $\theta_y$ between ±0.5 mrad is extracted, thus permitting the off-axis redshift (at maximum 35 eV) to be neglected. The wavelength dependent absorption of X-rays during transport is corrected for in the resulting spectrum.

This spectrum is shown in Figure 7(a), along with the probability model spectrum using $\kappa$=0.75, smoothed by including resolution-limiting effects, *i.e.* the estimated homogeneous broadening effects in the ICS spectrum as well as the calibrated resolution in the measurement system deduced from Figure 2. The measurement system rms broadening dominates at 260 eV; the contributions due to the angles in the electron beam, and to the finite laser pulse length are both ~10 eV, respectively. Also given in Figure 7(b) are the predictions of the L-W model, obtained using measured laser and electron beam parameters, with $\sigma_e$=40, and $\sigma_l$=30 $\mu$m. The L-W calculation, which does not suffer from weak divergence spectral artifacts, directly omits the measurement-specific spectral effects. As such, the measurement resolution effects must be convolved with the L-W numerical results, with the line shape also shown in Figure 7(b).

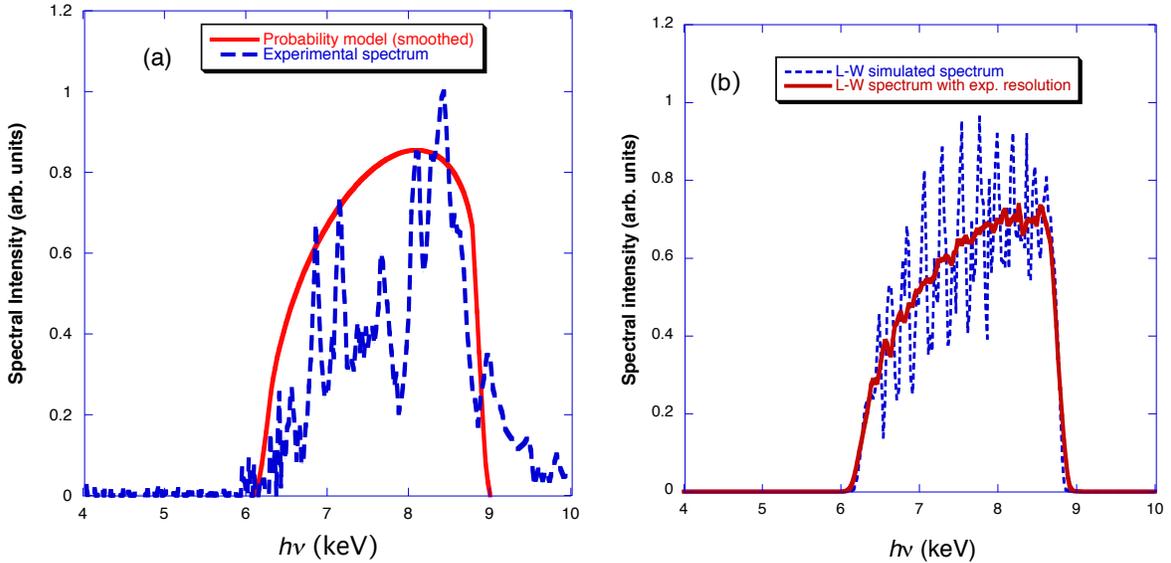

Figure 7. (a) Observed near-axis radiation spectrum extracted from DDS image in Figure 3(b) (blue dashed line). Also shown, predictions of probability model assuming $U_b$=65 MeV, $\kappa$=0.75, $a_0=1$ (red solid line, with predications smoothed to reflect determined gaussian resolution, per Figure 2). (b) Simulated spectrum from L-W calculation with same parameters, showing self-interference effects. Also shown, L-W spectrum smoothed for experimental resolution.

As expected, the observed spectrum extends from above 8.7 keV to <6 keV, due to the large maximum relative redshift $\Delta\lambda_{max}/\lambda_{r0} = \frac{1}{2}a_0^2 = 0.5$, or 0.33 in terms of relative photon energies. Further, the probability model and experimental data have similar spectral shape, descending from a larger spectral intensity near the smallest redshift to slightly lower intensity near the largest redshift, and having similar upper and lower edges — as previously noted in the spectral edge determinations in Figure 4. We note that the experimentally determined value of $\kappa$ reproduces the shape of the spectrum fairly well; choices of $\kappa$ more than 20% deviating from the nominal value of 0.75 do not give qualitative agreement in the aspects spectral shape discussed above. Other shots also display similar spectral shapes including the local fluctuation peaks seen in Figure 7(a).

The simulated L-W spectrum in Figure 7(b) also displays quasi-periodic peaks; there are associated with self-interference effects. It is interesting to ask whether these are related to those found in the experimental spectra. Indeed, Fourier analyses of both the experimental spectra and L-W predictions yield similar characteristics, displaying oscillations with approximately 0.3 keV periods each (or ~10-12 peaks inside of the spectral distributions). In the theoretical treatment, the scale of this period is set by the interference condition, which is a function maximum nonlinear redshift and the laser pulse length [22]. The measured, and computationally predicted, period is only a bit larger than the experimentally determined resolution of the curved multi-layer, $\sigma_E$=0.26 keV, however. To show the likelihood that the interference effect would be detected, the spectrometer resolution is convolved with the L-W simulation predictions, as shown in Figure 6(b). The oscillations in the spectrum are seen to barely survive in this simulation of the measurement. Thus this aspect of the nonlinear ICS physics was not likely to be observed in the experiment as it was performed. Any apparent oscillations are likely due despite the granular nature of the obtained spectrum arising from the finite number of photons detected MCP and associated screen, a phenomenon displayed in Figure 4. The statistical sampling of the data has been improved by taking the maximum width of the DDS in angle consistent with being able to ignore the off-axis redshift, but there are still notable statistical fluctuations in the data, as seen in Figure 7(a).

**CONCLUSIONS**

The results of measurements obtained from use of the curved multi-layer spectrometer described here thus clearly show, on a single-shot basis, compelling details of the double-differential ICS spectrum in the highly nonlinear regime where $a_0$ reaches unity. The analysis of these measurements agree well with the convoluted scattering probability function, illustrating clearly the red shifting effects associated with off-axis emission and nonlinear electron motion in the large laser field in the fundamental emitted band. Further, the shape and width of the ICS near-axis photon energy spectrum is seen to give insight into the relative laser-electron beam spots $\kappa$, as well as the degree of nonlinearity parameterized by $a_0$. The methodology employed here, as well as the push to considerably higher laser intensities (~2.5 times larger than those of Ref. 16), thus provides a much more detailed look into the electrodynamic processes involved in nonlinear inverse Compton scattering than previously obtained. As such, this investigation has, importantly, introduced a powerful new experimental method for understanding the intense laser-relativistic electron beam interaction, and also in the process also shed considerable light on the spectral structure of ICS in this limit, which is extremely relevant to applications. This method may be further improved to allow detection of self-interference effects in ICS, phenomena that have been inconclusively resolved by use of the bent crystal dispersive system. To investigate these effects, one may optimize the design of the spectrometer to improve the spectral resolution [25,26]. This improvement can be accomplished by using thinner, more numerous multi-layer strata with attendant change in incident angle. It will also be of interest, in the highly nonlinear limit, to examine the double differential spectrum associated with ICS higher harmonics in strong laser field, $a_L$~1 limit. Such issues may be taken up in future experiments that will seek to fully exploit the methods used in the present work.

This work supported by the US DOE contract DE-SC0009914, and US Dept. of Homeland Security Grant 2014-DN-077-ARI084-01.